\documentclass{aa}
\usepackage{graphicx}
\usepackage{txfonts}
\usepackage{subfigure}
\usepackage{hyperref}
\usepackage{orcidlink}

\begin{document} 

\title{Temporal evolution of the circumstellar disk orientation in the transient X-ray pulsar \mbox{GRO~J1008$-$57}}  
\titlerunning{Circumstellar disk orientation in the X-ray pulsar \mbox{GRO~J1008$-$57}}

\author{
Yongfeng Hu\inst{\ref{in:UTU}}\orcidlink{0009-0009-7560-9084}\thanks{E-mail: yongfeng.hu@utu.fi},
Hua~Xiao \inst{\ref{in:UTU}}\orcidlink{0009-0004-1288-4912},
Sergey~S.~Tsygankov \inst{\ref{in:UTU}, \ref{in:Tub},\ref{in:IHEP}}\orcidlink{0000-0002-9679-0793},
Long~Ji \inst{\ref{in:sysu}}\orcidlink{0000-0001-9599-7285},
Juri~Poutanen \inst{\ref{in:UTU}}\orcidlink{0000-0002-0983-0049},
Runting Huang\inst{\ref{Independent Researcher}}
}
          
\institute{Department of Physics and Astronomy, FI-20014 University of Turku,  Finland \label{in:UTU} 
\and
Institut f\"ur Astronomie und Astrophysik, Universit\"at T\"ubingen, Sand 1, D-72076 T\"ubingen, Germany \label{in:Tub} 
\and 
Key Laboratory of Particle Astrophysics, Institute of High Energy Physics, Chinese Academy of Sciences, Beijing 100049, China\label{in:IHEP} 
\and
School of Physics and Astronomy, Sun Yat-sen University, Zhuhai, 519082, People’s Republic of China \label{in:sysu}
\and
Independent Researcher, Guangdong, People’s Republic of China \label{Independent Researcher}
}

\authorrunning{Hu et al.}

\abstract 
{The transient X-ray pulsar GRO~J1008$-$57 was previously found to exhibit Type I outbursts occurring at stable orbital phases before its first observed Type II outburst in 2012. In this work, we extend the study to investigate the phase evolution after several Type II outbursts using long-term Swift/BAT and MAXI/GSC observations.
Our results reveal the orbital phases of Type I outbursts follow a step-like evolution: they remain largely stable over many orbital periods but undergo abrupt, small-amplitude jumps coincident with each Type II outburst. Such a step-like behavior is difficult to explain with the commonly proposed mechanisms involving a highly eccentric or precessing disk around Be star. The energetics of Type I X-ray outbursts shows a systematic increase before Type II outbursts, followed by a rapid decline and a subsequent gradual recovery. This behavior suggests cycles of disk depletion and reconstruction driven by Type II outbursts. 
Considering the small amplitude of each phase jumps, we propose that this step-like phase evolution may be related to the long orbital period of GRO~J1008$-$57, implying infrequent neutron star–disk interactions. After disk depletion by Type II outbursts, the disk around Be star has enough time to rebuild its density and restore geometric structure similar to its pre–Type II outburst state. Consequently, the orbital phases of subsequent Type I outbursts not only change very slightly but also can remain stable over many orbital periods until the next Type II-driven disk reconfiguration, yielding the observed step-like evolution.
}

\keywords{accretion, accretion disks -- ephemerides -- pulsars: individual: GRO~J1008$-$57 -- stars: emission-line, Be -- stars: neutron -- X-rays: binaries}

\maketitle

\section{Introduction} 
\label{Introduction}

Be/X-ray binaries (BeXRBs) are a common type of high mass X-ray binaries (HMXB) that contain a compact object (typically a neutron star) and a main-sequence O/B star (see \citealt{Reig2011Ap&SS.332....1R} for a review).
The ``Be'' designation of the O/B star originates from the characteristic Balmer-series emission lines observed in its optical spectrum.
It also exhibits an infrared excess relative to normal O/B stars \citep{Porter2003PASP..115.1153P,Rivinius2019IAUS..346..105R}.
These two observational properties are interpreted as clear indicators of a geometrically thin, circumstellar “decretion” disk that surrounds the Be star \citep{Porter2003PASP..115.1153P}.
This disk is believed to be formed by the high-speed rotation of the Be star that triggers the outflow of material from its equatorial plane, which then undergoes viscous diffusion into a Keplerian disk  \citep{Lee1991MNRAS.250..432L}. 
If the decretion disk expands beyond the Roche lobe, the neutron star can capture and accrete material from the disk.

The accreted matter is channeled along the magnetic field onto the magnetic poles of the neutron star, ultimately converting the gravitational energy of the accreting material into X-rays  \citep[see, e.g.,][]{Mushtukov_Tsygankov_2024_Review}. 
Most BeXRBs are X-ray transients and their behavior is characterized by two types of outbursting activity \citep{Negueruela1998A&A...336..251N}.
Type I X-ray outbursts (normal outbursts), which last for only a small fraction of the orbital period in long-period systems, are periodic in nature and are often observed at the periastron passage with a typical luminosity of $L_{\mathrm{x}} \sim 10^{36}$--$10^{37}$\,erg\,s$^{-1}$.
Type II X-ray outbursts (giant outbursts) occur much less frequently but have a larger luminosity of $L_{\mathrm{x}} > 10^{37}$\,erg\,s$^{-1}$ and last for a large fraction of an orbital period or even for several orbital periods, showing no preferred orbital phase \citep{Reig2011Ap&SS.332....1R, Martin2014ApJ...790L..34M,Okazaki2001A&A...377..161O}. 
Type~I outbursts are commonly interpreted to be the result of interactions that occur when the neutron star makes its closest approach to the circumstellar disk of the Be star in each orbital cycle \citep{Okazaki2001A&A...377..161O,Okazaki2013PASJ...65...41O}.
Type~II outbursts are believed to be related to a highly misaligned and increasingly eccentric decretion disk, which causes a prolonged interaction with the neutron star \citep{Okazaki2013PASJ...65...41O,Martin2014ApJ...790L..34M,Monageng2017MNRAS.464..572M}.

However, from the actual observations it has been discovered that the orbital phase of Type I outbursts does not always coincide with the periastron, and the outbursts may occur  either before or after the periastron. 
Moreover, during Type II outbursts, the phase of Type I outbursts may suddenly shift with the following gradual recovery over several hundred days.
Sometimes, the orbital phases of Type~I outbursts exhibit a periodic alternation between phase delays and phase advances. 
Detailed studies of the orbital phase variations of Type I outbursts in EXO 2030+375 have been reported by \cite{Wilson2002ApJ,Wilson2005ApJ,Wilson2008ApJ}, \cite{Baykal2008A&A...479..301B}, \cite{Laplace2017A&A...597A.124L}, and \cite{Huang2025ApJ...984...66H}, providing a representative example.

It should be emphasized that both Type I and Type II outbursts essentially originate from the interaction between the neutron star and the Be disk. 
Consequently, these puzzling orbital phase variations actually reflect the structure, the physical properties, and temporal evolution of the Be disk, which sometimes undergoes continuous changes.
Although there are some theoretical models describing these changes, the actual physical interaction between a neutron star and a Be disk remains difficult to study. 
Nevertheless, studying these phase variations in a larger sample of sources is essential from a phenomenological and physical point of view and provides key input for refining current theoretical models. 
However, this requires observations of sources that exhibit multiple and regular outbursts, and one of the most promising candidates for such a study is GRO~J1008$-$57.

GRO~J1008$-$57 is a BeXRB composed of a transient X-ray pulsar with a spin period of 93.5~s \citep{Stollberg1993IAUC.5836....1S}
and a B0e star at a distance of 3.6~kpc \citep{Fortin2022A&A}. 
The binary has an orbit with a period of 249.48~d and an eccentricity of 0.68 \citep{Coe2007MNRAS.378.1427C,Kuhnel2013A&A...555A..95K}. 
Since its discovery by the \textit{Compton Gamma Ray Observatory} (\textit{CGRO}) during the 1993 X-ray outburst  \citep{Stollberg1993IAUC.5836....1S,Wilson1994AIPC..308..451W}, it exhibited periodic Type~I outbursts around the periastron passage over many years.
Previously, \cite{Kuhnel2013A&A...555A..95K} reported that the orbital phase of Type~I outbursts in GRO~J1008$-$57 was remarkably stable between 2005 and 2012, before it exhibited the first Type II outburst.
However, since the works by \cite{Kuhnel2013A&A...555A..95K}, GRO~J1008$-$57 has experienced a total of four Type II outbursts, which were detected in November 2012 \citep{Kuehnel2012ATel.4577....1K}, January 2015 \citep{Kretschmar2015ATel.6917....1K}, August 2017 \citep{Sguera2017ATel10626....1S}, and June 2020 \citep{Nakajima2020ATel13750....1N,Wang2021JHEAp..30....1W}.
Based on these findings, therefore, GRO~J1008$-$57 is an ideal object to study the orbital phase evolution of  Type~I and Type~II outbursts. 
In this paper, we aim to investigate the orbital phase evolution of Type~I outbursts after several Type~II outbursts in GRO J1008$-$57 and compare it with previous studies. 
The paper is organized as follows. 
Section~\ref{Observations} briefly describes the observations. 
Data analysis and results are given in Sect.~\ref{Data Analysis and Results}. 
Finally, we discuss the results and present the conclusions in Sect.~\ref{sec:disc}.

\section{Observations} 
\label{Observations}

The Burst Alert Telescope (BAT) is one of the three instruments on board the \textit{Neil Gehrels Swift Observatory} \citep[\textit{Swift};][]{Gehrels2004ApJ} spacecraft designed to study gamma-ray bursts. Operating in the 15--150~keV energy range, the BAT provides $\sim$7 keV energy resolution, a sensitivity of $\sim10^{-8}$~erg~s$^{-1}$~cm$^{-2}$, and a 1.4~sr field of view. 
It  performs an all-sky hard X-ray survey with a sensitivity of $\sim$2~mCrab (systematic limit), while serving as a hard X-ray transient monitor \citep{Gehrels2004ApJ,Barthelmy2005SSR}. 
In order to estimate the flux variability during outbursts, we utilized daily light curves of GRO~J1008$-$57 provided by the Swift/BAT Hard X-ray Transient Monitor website.\footnote{\url{https://Swift/BAT.gsfc.nasa.gov/results/transients/GROJ1008-57}}

The Monitor of All-sky X-ray Image (MAXI; 2--30 keV) is  mounted at the International Space Station). 
Its scientific instruments consist of two types of X-ray cameras with a wide field-of-view: the Solid-state Slit Camera (SSC) and the Gas Slit Camera (GSC). 
The main role of the GSC is to serve as the X-ray all-sky monitor \citep{Matsuoka2009PASJ}.
We also tracked the outburst activity of GRO~J1008$-$57 in the soft X-ray band using data from the MAXI/GSC website.\footnote{\url{https://MAXI/GSC.riken.jp/star_data/J1009-582/J1009-582.html}}

To study total energy release during Type~I outbursts, the Swift/BAT  and MAXI/GSC  count rates were converted to energy fluxes assuming the Crab Nebula as a standard candle. 
We adopted a conversion  $1~\mathrm{Crab}=0.22~\mathrm{count~s^{-1}~cm^{-2}}=1.4\times10^{-8}~\mathrm{erg~cm^{-2}~s^{-1}}$ for the BAT data in the 15--50~keV band \citep{Krimm2013ApJS} and $1~\mathrm{Crab}=3.6~\mathrm{count~s^{-1}~cm^{-2}}=3.1\times10^{-8}~\mathrm{erg~cm^{-2}~s^{-1}}$ for the MAXI data in the 2--20 keV band \citep{Matsuoka2009PASJ}.
The count rates were converted to energy fluxes and used to estimate the total radiated energy during each outburst assuming a source distance of 3.6~kpc.

\section{Data Analysis and Results} 
\label{Data Analysis and Results}

\subsection{Light curve} 
\label{Light curve}

\begin{figure*}
\centering
\includegraphics[width=5.9in]{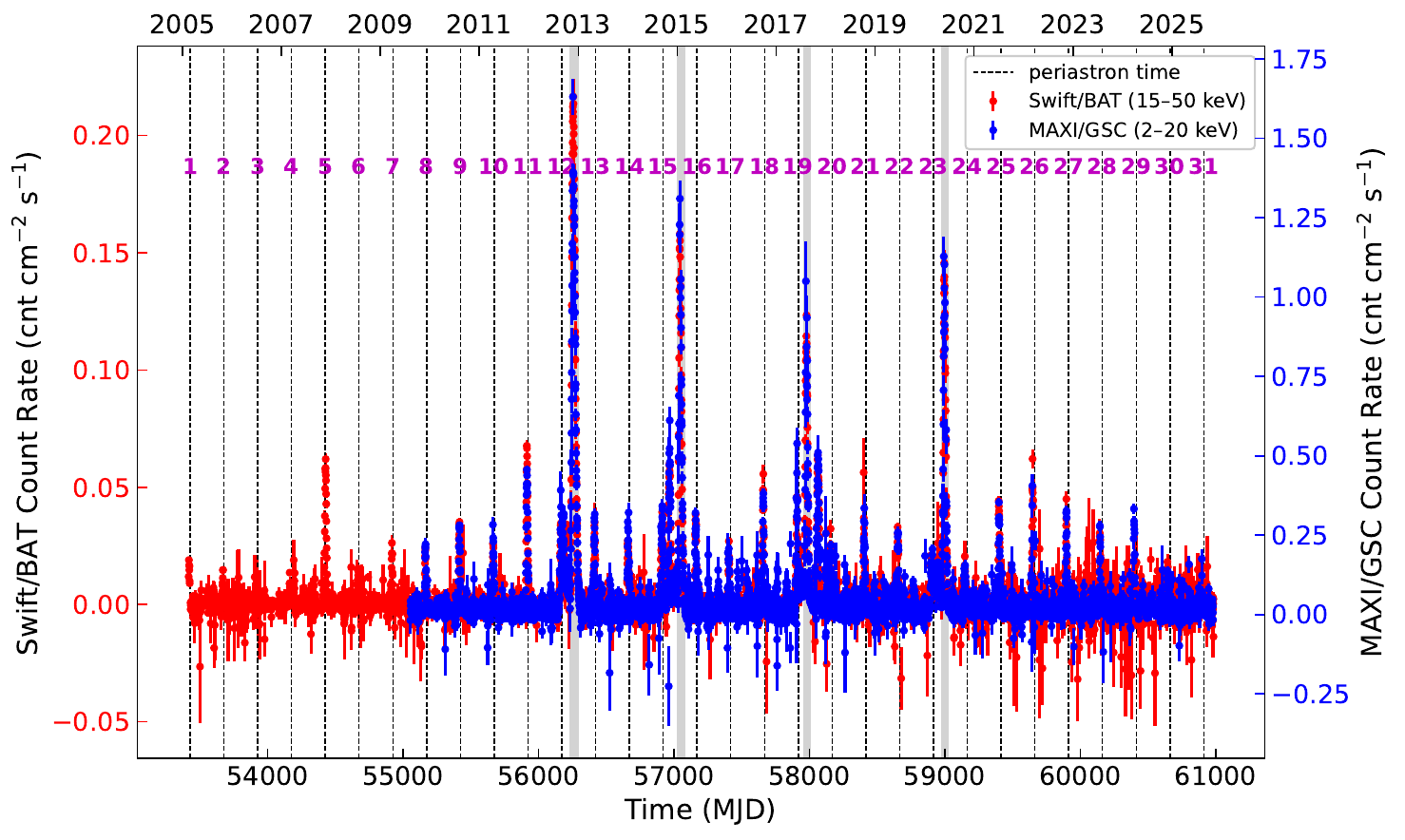}
\caption{Long-term light curves of GRO~J1008$-$57 obtained with \textit{Swift}/BAT (15--50 keV; red points) and MAXI/GSC (2--20 keV; blue points). The shaded regions indicate the four Type II outbursts occurring in 2012, 2015, 2017, and 2020.
The vertical dashed lines indicate the periastron times of each orbital period, calculated using the ephemeris from \cite{Kuhnel2013A&A...555A..95K}. The purple numbers represent the 31 TypeI outbursts detected by \textit{Swift}/BAT and the 24 detected by MAXI/GSC, respectively. The profiles of all Type~I outbursts are shown in Fig.~\ref{fig:evolution}. 
}
\label{fig:counts}
\end{figure*}

Figure~\ref{fig:counts} shows the light curve of GRO~J1008$-$57 as observed by \textit{Swift}/BAT and MAXI/GSC, revealing regular and stable luminosity variations of the source over long timescales.
Since Type~I outbursts are generally believed to occur near periastron passage, we therefore calculated the times of periastron passage for each orbital period using the ephemeris from \cite{Kuhnel2013A&A...555A..95K}, which provides an orbital period $P=249.48$\,d and a periastron time MJD~54424.71. 
A total of 31 and 24 Type I outbursts can be identified in the \textit{Swift}/BAT and MAXI/GSC observations, respectively. 
In our numbering convention, the epoch of the Type I outburst number $n$ can be expressed as its periastron time $T_{n}$ = MJD 53426.789 + $n\times$249.48, where $n$ is the number of orbits since the outburst in 2005. 
In addition, the positions of four Type~II outbursts 
are marked with gray shaded regions, allowing us to specifically examine the phase variations of Type~I outbursts occurring in their vicinity.

\subsection{The start and peak times of Type I outbursts} 
\label{sec:start}

\begin{figure*}
\centering
\includegraphics[width=0.27\linewidth]{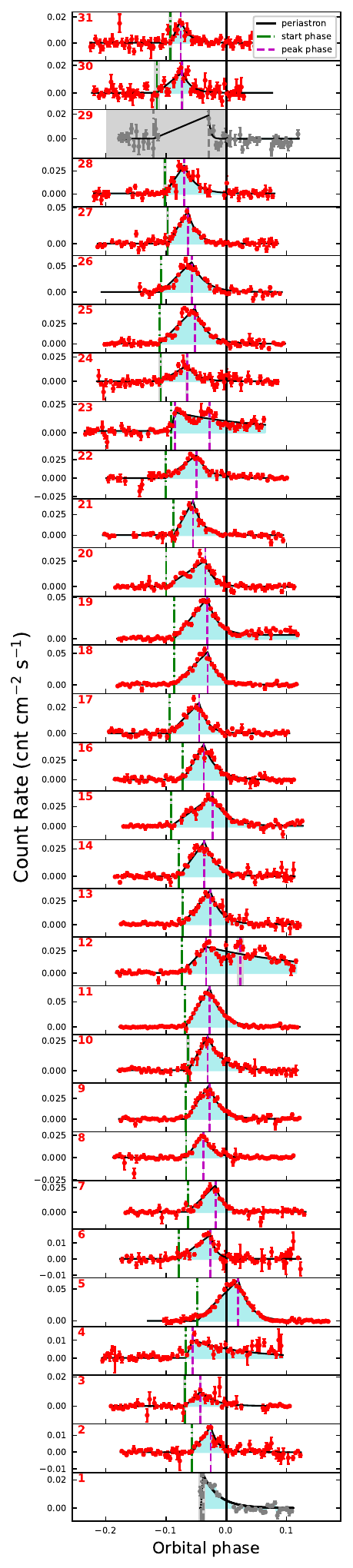}
\includegraphics[width=0.27\linewidth]{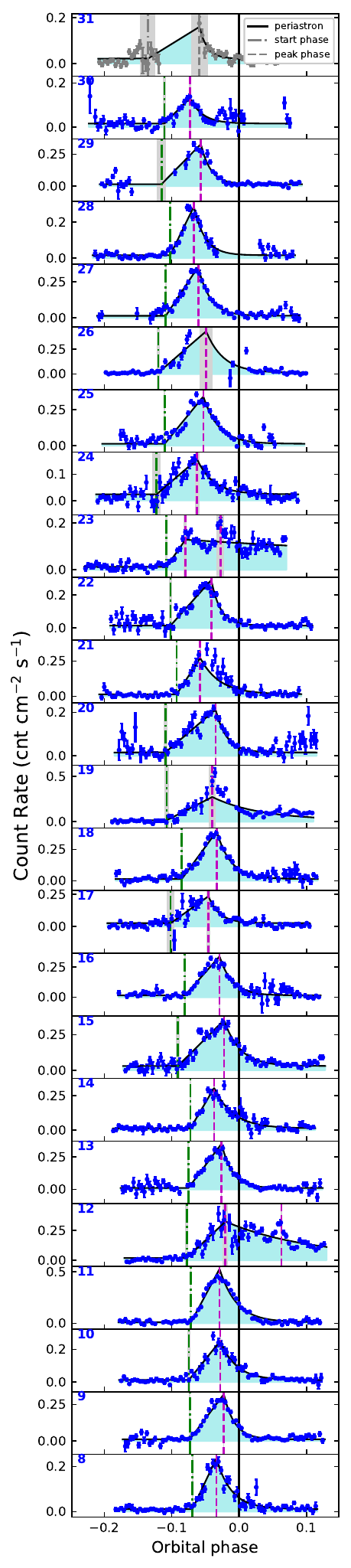}
\includegraphics[width=0.395\linewidth]{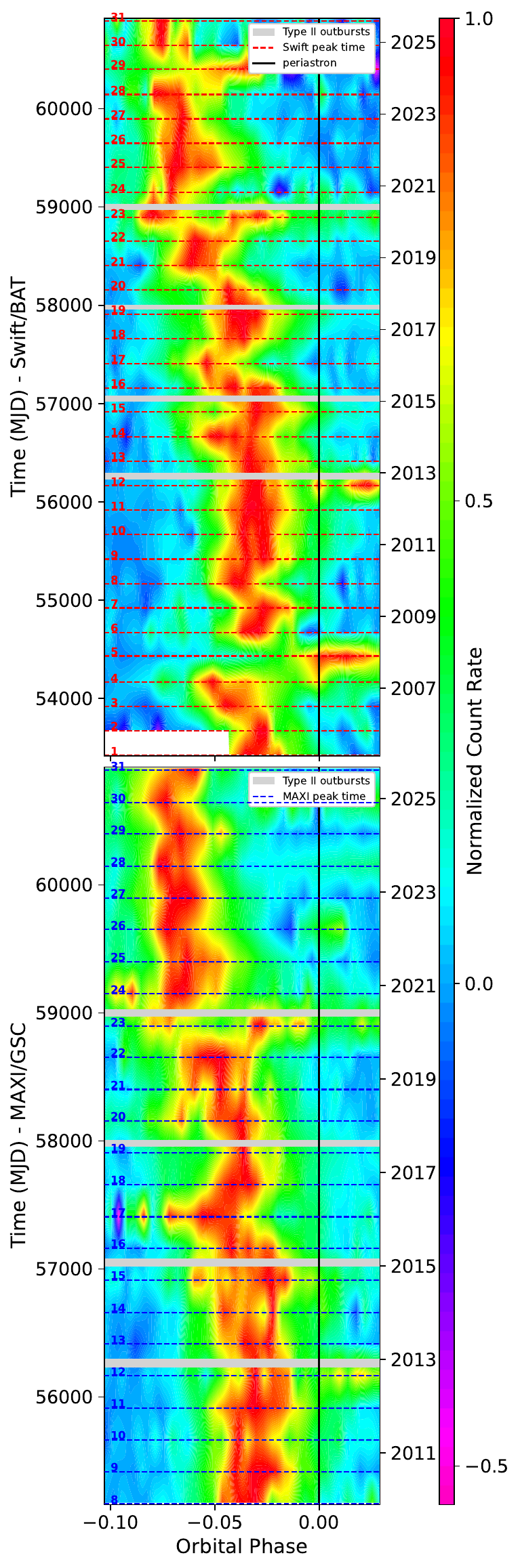}
\caption{Evolution of outburst profiles of GRO~J1008$-$57 .
Left panel: Profiles of Type I outbursts observed with \textit{Swift}/BAT, arranged from bottom to top in order of occurrence time. The green and purple vertical dashed lines indicate the burst onset times and peak times, respectively, while the black vertical line marks the periastron at orbital phase 0. 
Bursts \#1, \#29 (\textit{Swift}/BAT), and \#31 (MAXI/GSC), shown in gray, are excluded from the analysis due to data gaps.
The cyan shaded regions used for calculating the total energy released (Fig. \ref{fig:flux}) during the Type I outbursts indicate that the burst profiles cover a range of $\pm0.15$ orbital period around the peak times.
Middle panel: Same as the left panel, but for the profiles of Type I outbursts observed with MAXI/GSC. Right panel:
Evolution of the color-coded intensity profiles of all Type I outbursts observed by \textit{Swift}/BAT and MAXI/GSC, with the peak intensity of each outburst normalized to unity. 
The horizontal dashed line and numbers indicate the peak times of each Type I outbursts, and the gray shaded regions show the Type II outbursts.}
\label{fig:evolution}
\end{figure*}

In order to determine when Type I outbursts occurred during each orbital period, we model the temporal profiles of Type I outbursts in GRO~J1008$-$57 using a simplified  fast-rise exponential-decay (FRED) function \citep[e.g.,][]{Rikame2025MNRAS}, comprising a linear rise, an exponential decay, and constant background component, as given by
\begin{equation}
F_{\mathrm{burst}}(t) =
\begin{cases}
C_1, & \text{if } t < t_{\rm start}, \\[6pt]
A \dfrac{t - t_{\rm start}}{t_{\rm peak} - t_{\rm start}} + C_1,
& \text{if } t_{\rm start} \le t < t_{\rm peak}, \\[10pt]
C_2 + (A + C_1 - C_2)\ {\rm e}^{-\frac{t - t_{\rm peak}}{\tau}},
& \text{if } t \ge t_{\rm peak},
\end{cases}
\label{eq:fred}
\end{equation}
where $t$ is the time, $t_{\rm start}$ and $t_{\rm peak}$ denote the start and peak times of the outburst, $A$ is the outburst amplitude, $\tau$ is the exponential decay timescale, 
the constants $C_1$ and $C_2$ represent the local persistent emission levels before and after the peak of the outburst, respectively.
This model enables us to simultaneously fit and estimate both the onset and peak times of all Type I outbursts in GRO~J1008$-$57.\footnote{We also applied the Gaussian model adopted by \cite{Laplace2017A&A...597A.124L} and \cite{Huang2025ApJ...984...66H} to determine the peak times, and obtained results similar to those derived from the FRED model.}
Outbursts \#1 and \#29 (\textit{Swift}) and outburst \#31 (MAXI), which have significant data gaps (Fig. \ref{fig:evolution}), were excluded from our sample.
The fitting time interval for each orbital period is set within $\pm0.5$ orbital periods around the periastron passage, and the uncertainties of the fitted start and peak times were estimated from the covariance matrix of the fit and correspond to $1\sigma$ confidence intervals.

\begin{figure*}
\centering
\includegraphics[width=0.45\linewidth]{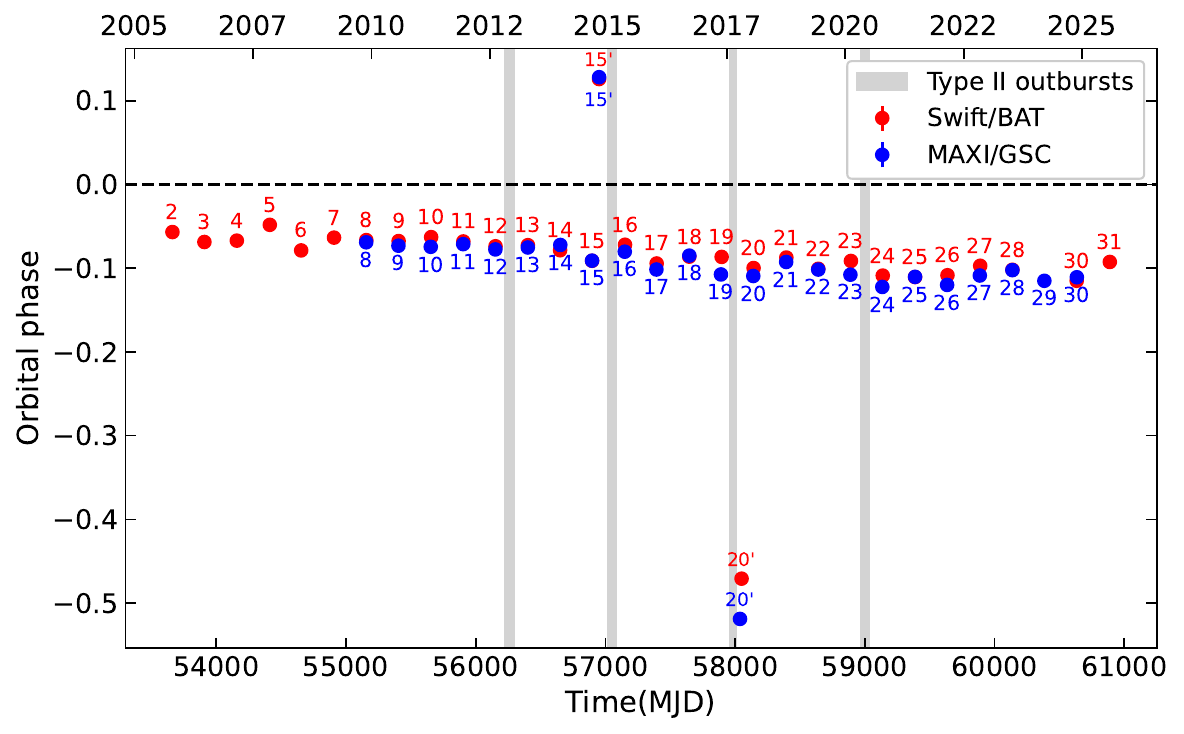}
\includegraphics[width=0.45\linewidth]{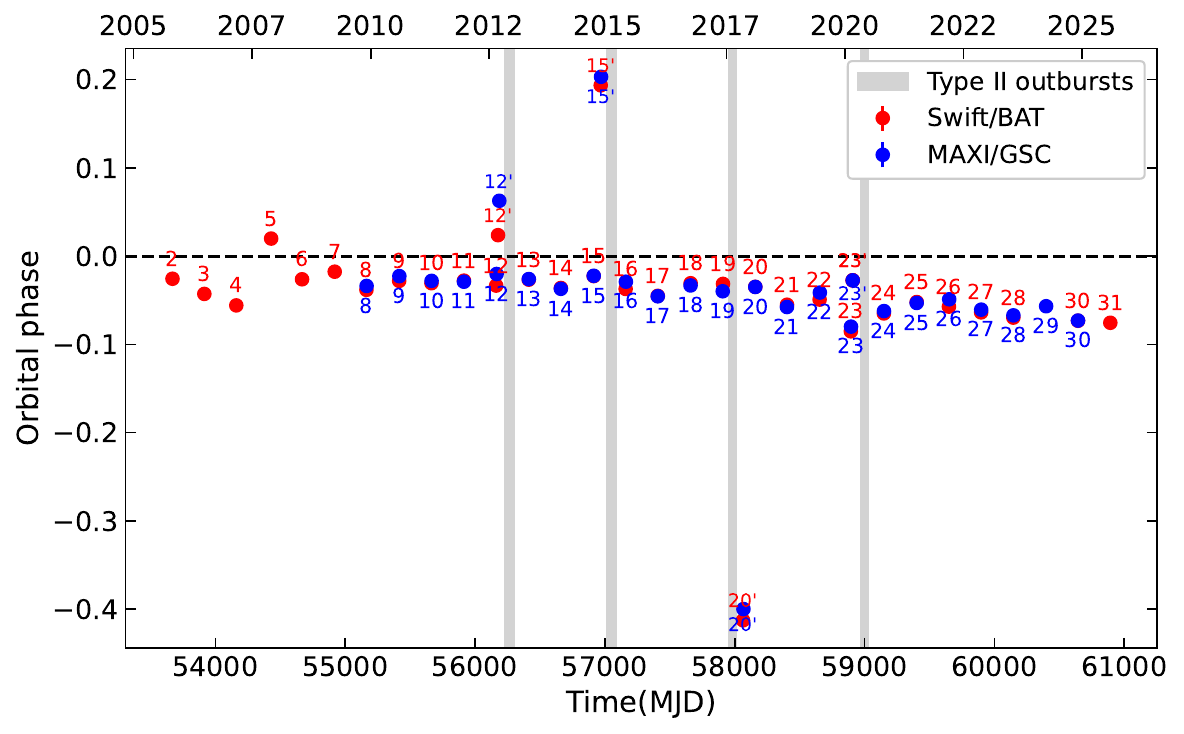}
\caption{Left panel: Orbital phases of the onset times of Type I outbursts from GRO~J1008$-$57 observed with \textit{Swift}/BAT (red numbers) and MAXI/GSC (blue numbers). 
The horizontal dashed line indicates the orbital phase 0, and the shaded vertical regions present the Type II outbursts. Right panel: Same as the left, but showing the orbital phases of the peak times of all Type I outbursts.}
\label{fig:total_start_peak}
\end{figure*}

After obtaining the best-fit timing results using the FRED model, we calculated the orbital phases of the onset and peak times of all Type I outbursts by computing their time differences relative to the corresponding periastron passages time, as shown in Fig.~\ref{fig:total_start_peak}.
The orbital phases of the fitted onset and peak times derived from the two instruments are generally consistent.

\begin{figure*}
\centering
\includegraphics[width=0.9\linewidth]{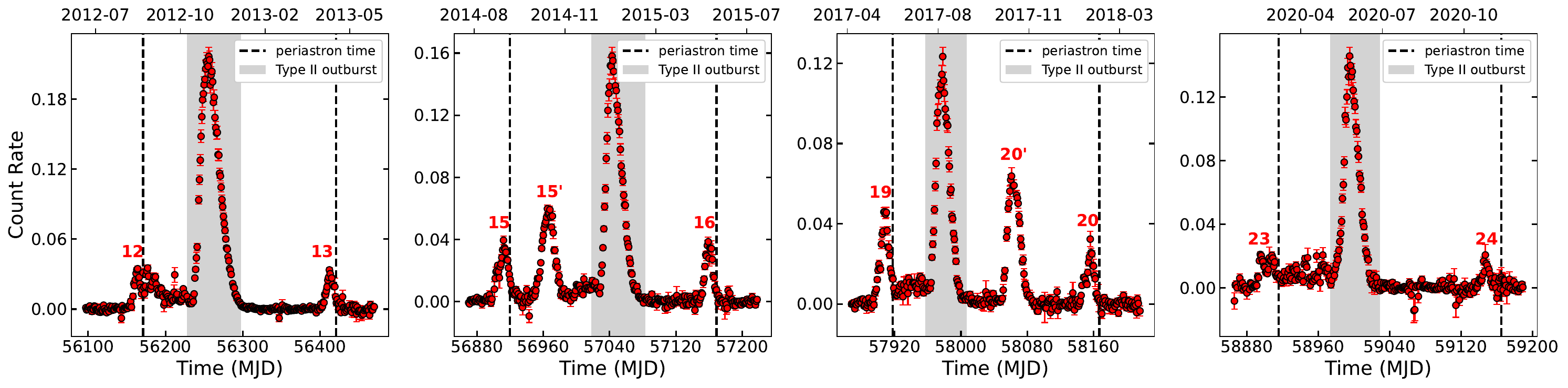}
\includegraphics[width=0.9\linewidth]{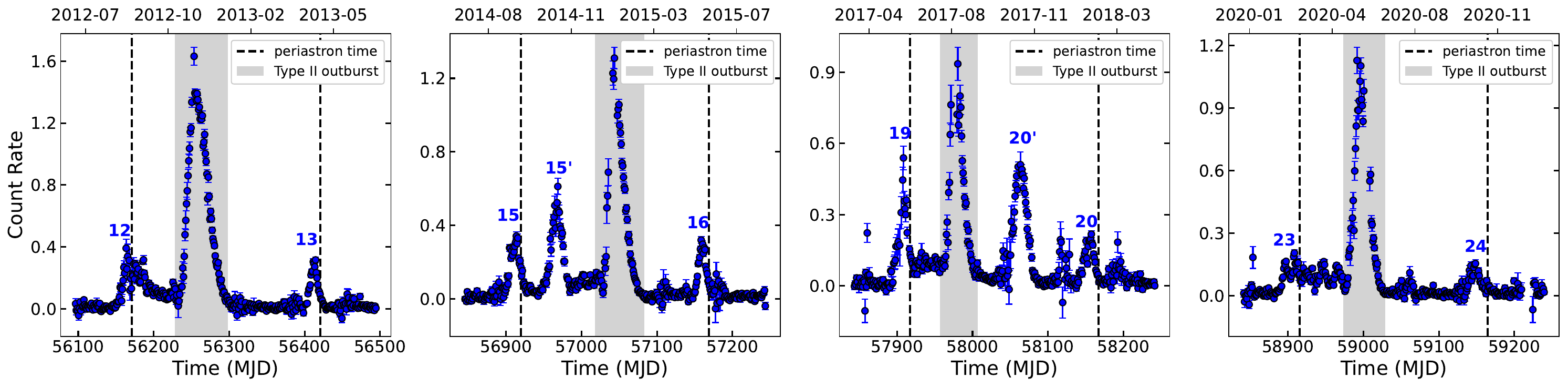} 
\caption{Light curves of GRO~J1008$-$57 close to the Type II outbursts from the \textit{Swift}/BAT (upper panels) and MAXI/GSC (lower panels) data. 
The vertical dashed lines present the periastron times, and the shaded regions indicate the Type II outbursts.
}
\label{fig:TypeIIburst}
\end{figure*}

Throughout the entire observational history of GRO~J1008$-$57, 
the orbital phases of both the onset and peak times for almost all bursts are consistently and periodically clustered in the vicinity of periastron, more specifically prior to the periastron passage.
However, Type I outbursts occurring close to the four Type II outbursts in 2012, 2015, 2017, and 2020 exhibit more complex behaviors, as illustrated by the detailed light-curve profiles shown in Fig.~\ref{fig:TypeIIburst}.
Near the 2012 and 2020 Type II outbursts, the profiles of bursts \#12 and \#23 exhibit multi-peaked structures (Fig.~\ref{fig:evolution}).
In contrast, near the 2015 and 2017 Type II outbursts, bursts \#$15'$ and \#$20'$ are regarded as additional outbursts (see Fig. \ref{fig:TypeIIburst}), exhibiting significant abrupt phase shifts relative to periastron.
However, the system subsequently returned rapidly to the normal periastron-associated Type I outburst pattern within one orbital period. 
For the same Type I outburst, the \textit{Swift}/BAT and MAXI/GSC observations not only yield consistent onset and peak times derived from the FRED model fitting but also show similar light-curve profiles.

\begin{figure*}
\centering
\includegraphics[width=0.45\linewidth]{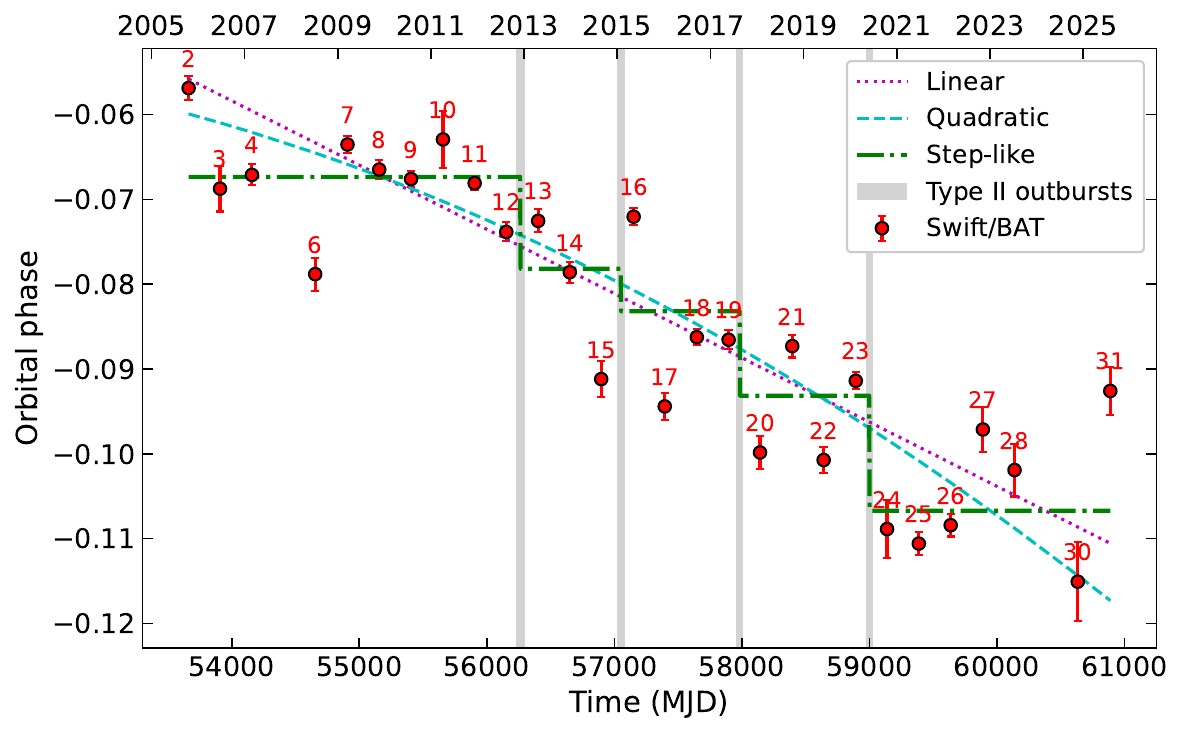}
\includegraphics[width=0.45\linewidth]{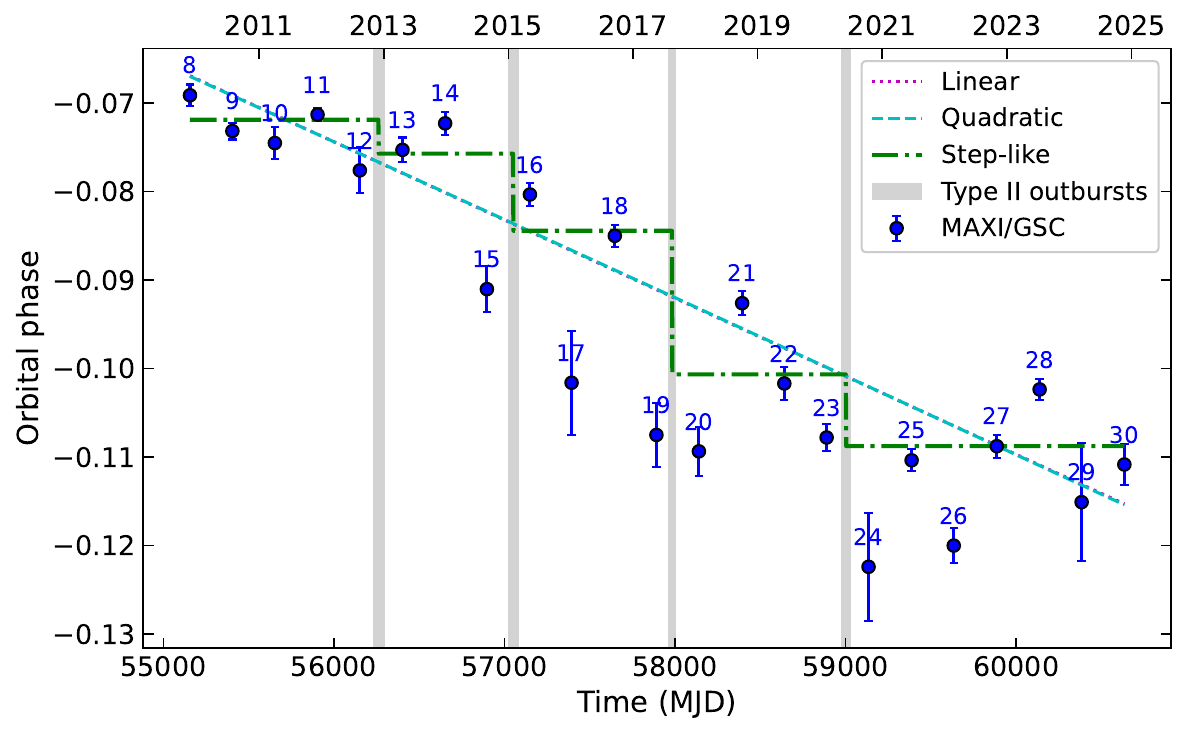}
\caption{Evolution of the orbital phase of the onset times of Type I outbursts in GRO~J1008$-$57 using \textit{Swift}/BAT (left panels) and MAXI/GSC (right panels) data and the best-fit models.  The shaded vertical regions indicate the Type II outbursts. The purple, cyan, and green dotted, dashed, and dash-dotted lines represent the linear, quadratic, and step-like models, respectively, with the step-like model providing the best fit to both the \textit{Swift}/BAT and MAXI/GSC data. The four step boundaries are fixed at the peak times of the four Type II outbursts.}
\label{fig:step_phase_fig}
\end{figure*}

\begin{table*} 
\centering
\caption{Step-like model parameters of the evolution of the orbital phase of the onset times of Type I outbursts in GRO~J1008$-$57 from the \textit{Swift}/BAT and MAXI/GSC data. }
\label{tab:step_phase_start}
\begin{tabular}{c c c c c c}
\hline\hline
Step & $t_{\rm start}$ & $t_{\rm end}$ 
     & Avg. phase 
     & $\Delta \phi$ 
     & $\Delta\phi/\Delta t$ (d$^{-1}$) \\
\hline
\multicolumn{6}{c}{{\textit{Swift}/BAT}} \\
1 & 53661 & 56263 
  & $-0.0673 \pm 0.0004$ 
  & -- 
  & -- \\

2 & 56263 & 57050 
  & $-0.0782 \pm 0.0008$ 
  & $-0.0108 \pm 0.0009$ 
  & $(-1.38 \pm 0.12)\times10^{-5}$ \\

3 & 57050 & 57982 
  & $-0.0832 \pm 0.0005$ 
  & $-0.0050 \pm 0.0010$ 
  & $(-5.37 \pm 1.08)\times10^{-6}$ \\

4 & 57982 & 59001 
  & $-0.0932 \pm 0.0007$ 
  & $-0.0100 \pm 0.0009$ 
  & $(-9.77 \pm 0.85)\times10^{-6}$ \\

5 & 59001 & 60889 
  & $-0.1067 \pm 0.0008$ 
  & $-0.0136 \pm 0.0010$ 
  & $(-7.18 \pm 0.55)\times10^{-6}$ \\

\hline
\multicolumn{6}{c}{{MAXI/GSC}} \\
1 & 55154 & 56263 
  & $-0.0719 \pm 0.0005$ 
  & -- 
  & -- \\

2 & 56263 & 57050 
  & $-0.0757 \pm 0.0009$ 
  & $-0.0038 \pm 0.0010$ 
  & $(-4.87 \pm 1.29)\times10^{-6}$ \\

3 & 57050 & 57982 
  & $-0.0844 \pm 0.0009$ 
  & $-0.0087 \pm 0.0012$ 
  & $(-9.35 \pm 1.32)\times10^{-6}$ \\

4 & 57982 & 59001 
  & $-0.1007 \pm 0.0009$ 
  & $-0.0162 \pm 0.0012$ 
  & $(-1.59 \pm 0.12)\times10^{-5}$ \\

5 & 59001 & 60635 
  & $-0.1088 \pm 0.0006$ 
  & $-0.0081 \pm 0.0011$ 
  & $(-4.96 \pm 0.06)\times10^{-6}$ \\

\hline
\end{tabular}
\tablefoot{Columns: 
(1) Step number; 
(2) $t_{\rm start}$, start time of the step (MJD); 
(3) $t_{\rm end}$, end time of the step (MJD); 
(4) Average orbital phase during the step; 
(5) $\Delta \phi$, phase change within the step; 
(6) $\Delta \phi / \Delta t$, average rate of phase evolution, where $\Delta t = t_{\rm end}-t_{\rm start}$.}
\end{table*}

\begin{figure*}
\centering
\includegraphics[width=5in]{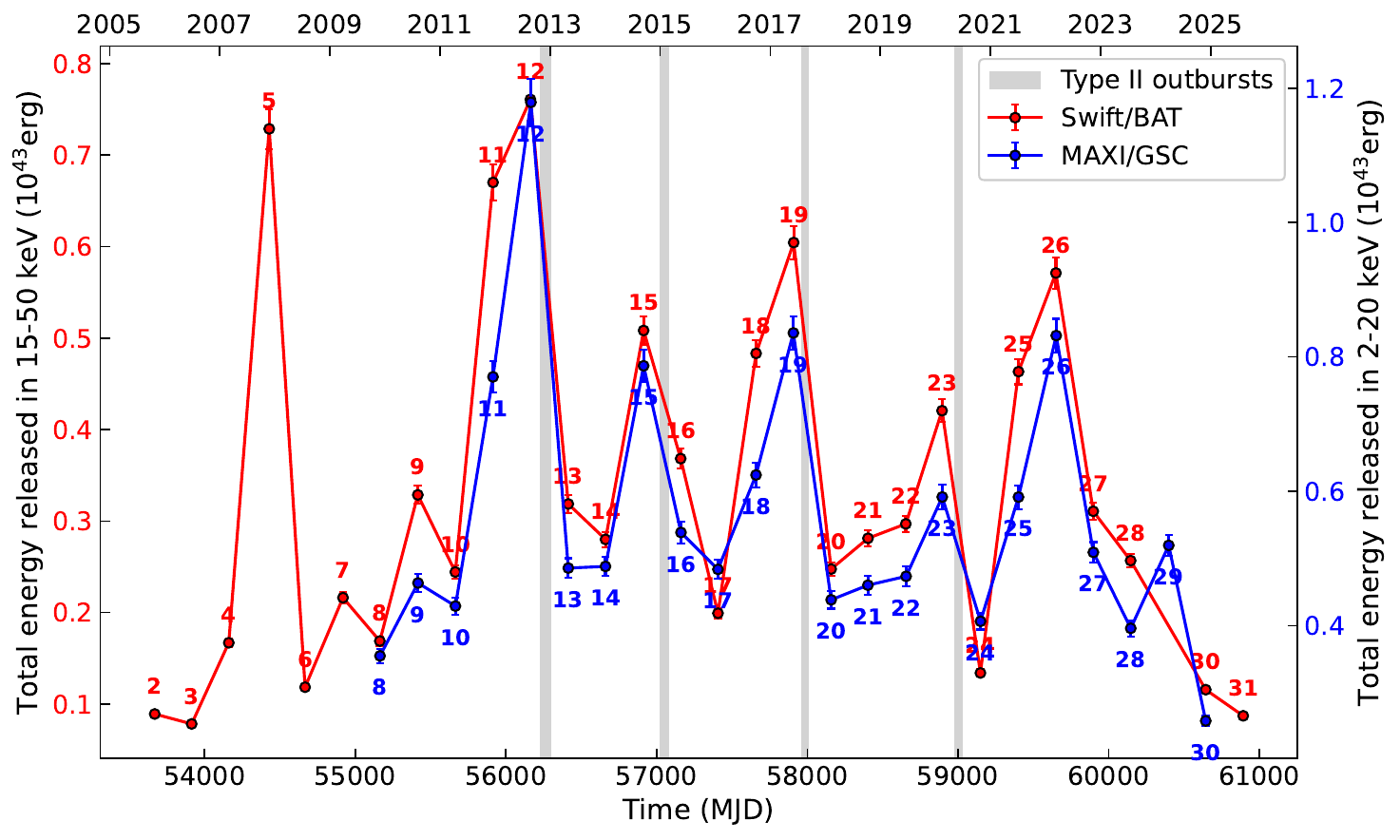}
\caption{Evolution of the total energy release of all Type I outbursts in GRO J1008$-$57, obtained by integrating the fitted model curve over a time interval of $\pm0.15$ orbital period around the burst peak (i.e., the area of the cyan shaded regions in Fig.~\ref{fig:evolution}). The red and blue points denote the \textit{Swift}/BAT and MAXI/GSC results, respectively. The shaded regions indicate the Type II outbursts.}
\label{fig:flux}
\end{figure*}

\subsection{Orbital phase evolution of the Type I outbursts} 
\label{sec:orb_evo} 

To further characterize the long-term orbital phase evolution of the Type I outbursts in GRO~J1008$-$57, we present the profiles and corresponding color-coded intensity map of all Type I outburst light curves observed by \textit{Swift}/BAT and MAXI/GSC in Fig.~\ref{fig:evolution}. 
Most Type I outbursts recur in a stable and periodic pattern close to, and typically before, periastron phase.
Previously, \cite{Kuhnel2013A&A...555A..95K} also analyzed the orbital phase evolution of Type I outbursts of GRO~J1008$-$57 observed with \textit{Swift}/BAT between 2005 and 2012 (i.e., bursts \#1--12, see Fig.~\ref{fig:total_start_peak}), all of which occurred before the first observed Type II outburst,  
and found that no phase evolution was present during this epoch, which is also in agreement with our results.
However, in our work, we note that the intensity map of the Type I outbursts shows that phase shifts begin to appear starting from burst \#13, which subsequently followed the first observed Type II outburst in 2012. The most significant phase shift is observed after the fourth Type II outburst in 2020.
To quantify this phase shift, we fitted the orbital phase evolution of the onset times of all the Type I outbursts with the linear ($\phi(t) = a + bt$), quadratic ($\phi(t) = a+bt+ct^2$), and step-like  model: 
\begin{equation}
\phi(t) = \left\{  \begin{array}{l} 
C_1, \ t < t_1, \\
C_2, \ t_1 \le t < t_2, \\
C_3, \ t_2 \le t < t_3, \\
C_4, \ t_3 \le t < t_4, \\
C_5, \ t \ge t_4,
\end{array}, 
\right. 
\end{equation}
where $t$ is the time, $t_1-t_4$ represent the boundary times of the step-like model, which are fixed at the peak times of the four Type II outbursts. 
Burst 5 (Swift/BAT), which exhibits a significant phase shift and unusually high luminosity (colored map in Figs.~\ref{fig:evolution} and \ref{fig:flux}), may represent an intermediate state between Type I and Type II outbursts. 
Adopting the same treatment as in \cite{Kuhnel2013A&A...555A..95K}, this burst is excluded from our fitting. The results of the fit are shown in Fig.~\ref{fig:step_phase_fig}.
Then we applied the Akaike Information Criterion (AIC=$\chi^{2}+2k+(2k^{2}+2k)/(n-k-1)$, where $\chi^2$ is the chi square value, $k$ is the number of free parameters, and $n$ is the number of data points) to choose the best model and found that the step-like model yields the lowest AIC value compared to both the linear and quadratic models for both \textit{Swift}/BAT and MAXI/GSC data. 
For the onset-phase fits, the $\Delta$AIC values between the step-like model and the linear and quadratic models are 76.9 and 37.9 for \textit{Swift}/BAT, and 81.7 and 84.3 for MAXI/GSC, respectively.
The corresponding chance improvement probability, calculated as $e^{-\Delta \text{AIC}/2}$ \citep{Konig2020A&A...643A.128K}, indicates that the significance of the difference between the step-like model and both the linear and quadratic models exceeds $5\sigma$.
This demonstrates that the step-like model provides the best fit to the orbital phase evolution of the onset times of the Type I outbursts in GRO~J1008$-$57 observed by  \textit{Swift}/BAT and MAXI/GSC. These results show that the orbital phase evolution does not follow a continuous long-term gradual drift as described by the linear and quadratic models.
Table~\ref{tab:step_phase_start} presents the best fit results of the step-like model for the \textit{Swift}/BAT and MAXI/GSC, it can be seen that all the phase jumps exhibit relatively small amplitudes.
Given that the peak phase depends on the outburst morphology (e.g., multi-peaked structures) and duration, we regard it as a less reliable tracer of the neutron star–disk interaction and therefore did not analyze its evolution.

We note that if the X-ray emission in GRO~J1008$-$57 is mediated by an accretion disk, the onset time of Type I outbursts may not directly reflect the instantaneous conditions in the Be circumstellar disk. As discussed by \cite{Tsygankov2017}, the source exhibits a moderate X-ray luminosity throughout the entire orbital cycle, which was interpreted in terms of different accretion regimes. As a result, the observed flux increase may occur with some delay associated with the transition of the accretion disk into the hot (ionized) state.
Nevertheless, relative shifts in the onset times of Type I outbursts from one orbital cycle to another can still serve as a tracer of the long-term evolution of the Be disk geometry.

Furthermore, we also investigated the long-term evolution of the total energetics (fluence) of Type I outbursts in GRO~J1008$-$57. The energy released in each outburst was estimated by integrating the fitted model curve over a time interval of $\pm0.15$ orbital phase around the outburst peak (Fig.~\ref{fig:flux}). 
The burst fluences inferred from \textit{Swift}/BAT and MAXI/GSC data are consistent with each other. 
Interestingly, in contrast to the orbital phases of Type I outburst onset, which transition between distinct piecewise stable states between Type II outbursts, their total energy release exhibits irregular long-term oscillatory variations. 
The behavior of the Type I outburst energetics generally shows a rise before the Type II outburst followed by a decline afterwards and a subsequent gradual recovery, although the overall picture may be more complicated. Similar rise-and-fall behaviour is also seen in outburst 5 and outburst 26 (i.e. before the first and after the last Type II outbursts), although they are not associated with Type II outbursts. In addition, the post-third-Type-II-outburst interval (2017--2020) shows a somewhat slower recovery in fluence compared to the other cases.

\section{Discussion and conclusion} 
\label{sec:disc}

In this paper, we investigate the temporal evolution of the orbital phase of Type I outbursts, with particular attention to the effects of several Type II outbursts, using observations from \textit{Swift}/BAT and MAXI/GSC. Our work substantially expands upon the findings of \cite{Kuhnel2013A&A...555A..95K}, who showed that the orbital phase of Type I outbursts in GRO~J1008$-$57 remained generally stable prior to the first observed Type II outburst in 2012.

In general, our study reveals that the phases of Type I outbursts of GRO~J1008$-$57 follow a step-like evolution, remaining stable over multiple orbital periods and exhibiting abrupt, small amplitude phase jumps (up to $-$0.0162 in orbital phase, see Table~\ref{tab:step_phase_start}) coincident with each Type II outbursts. 
Each jump presumably occurs over a short timescale (tens of days) due to the relatively brief duration of Type II outbursts.
We note that the orbital period reported by \cite{Kuhnel2013A&A...555A..95K} has an uncertainty of $\pm 0.04$ d. Over the observed $\sim$30 orbital cycles, this could accumulate to an uncertainty of $\sim$1.2 d in the predicted periastron time. However, this value is much smaller than the observed total phase shift of $\sim$0.04 (corresponding to $\sim$10~d) and therefore cannot explain the observed effect.

However, detailed studies of another transient Be/XRP EXO~2030+375 reveal distinctly different evolution features. In that system, phase shifts are not only triggered by Type II outbursts, but are sometimes also associated with Type I outbursts \citep{Laplace2017A&A...597A.124L,Huang2025ApJ...984...66H}. 
It exhibits a potential 5--10~yr periodicity alternation between positive phase shifts associated with Type II outbursts and negative phase shifts associated with Type I outbursts. 
Each phase shift showed a large amplitude (up to 0.5 in orbital phase) and was followed by long-term (hundreds of days) phase variations or a gradual recovery towards the periastron phase \citep{Wilson2002ApJ,Wilson2005ApJ,Wilson2008ApJ,Baykal2008A&A...479..301B,Laplace2017A&A...597A.124L,Huang2025ApJ...984...66H}.

In order to explain the periodic positive and negative phase alternation shifts in EXO 2030+375,
\cite{Laplace2017A&A...597A.124L} applied the Kozai-Lidov (KL) framework investigated by \cite{Martin2014ApJ...792L..33M} and \cite{Fu2015ApJ...807...75F,Fu2015ApJ...813..105F}, in which a long-term periodic exchange occurs between the eccentricity and inclination of the Be disk, and calculated an KL oscillation timescale of about 10 years for EXO~2030+375, remarkably matching its observed phase alternation shift period as above.

Using the method described in \cite{Martin2014ApJ...792L..33M} and \cite{Laplace2017A&A...597A.124L}, 
assuming a rigid Be disk of radius $R$ with a surface density profile following a power law, $\Sigma \propto R^{-p}$, 
the KL oscillation timescale, $\tau_{\rm KL}$, can be estimated as
\begin{equation}
\frac{\tau_{\textrm{KL}}}{P_{\textrm{orb}}}\approx \frac{\left(4-p\right)}{\left(\frac{5}{2}-p\right)}\left({1-e}\right)^{-1}\sqrt{\frac{M_{\textrm{Be}}}{M_{\textrm{NS}}}\left(\frac{M_{\textrm{Be}}}{M_{\textrm{NS}}} + 1\right)}\textrm{ ,}
\end{equation}
where $M_{\rm Be}$ and $M_{\rm NS}$ denote the masses of the Be star and the neutron star, respectively, $e$ is the orbital eccentricity, and $P_{\rm orb}$ is the orbital period of the binary system. 
For the unknown parameters, we adopt typical values of $p=1.5$ \citep{Martin2014ApJ...792L..33M}, $M_{\rm Be}=20\,M_\odot$, and $M_{\rm NS}=1.4\,M_\odot$ \citep{Okazaki2001A&A...377..161O}.
Using the orbital parameters described in \cite{Kuhnel2013A&A...555A..95K}, we also estimate a KL oscillation timescale of about 78.9~yr for GRO~J1008$-$57. 
This timescale far exceeds the available observational history of GRO~J1008$-$57, therefore the KL-driven mechanism cannot be excluded for the source.
From another side, this mechanism is expected to produce substantial eccentricity growth and large phase shifts, as observed in EXO 2030+375, but not seen in GRO~J1008$-$57, thereby putting KL-driven phase shifts under question.

To explain long-term phase shift in EXO~2030+375, such as the phase shifts associated with Type I outbursts spanning MJD 50000--50600, 
\cite{Wilson2002ApJ} proposed that these phase shifts were caused by a density perturbation propagating through the Be disk, i.e., a global one-armed oscillation, which produces a non-axisymmetric disk and can shift the outburst phase when the neutron star interacts with the perturbed region. 
This process operates on long timescales, with a characteristic period of 15$\pm$3~yr for EXO~2030+375. 
Such density perturbations have been inferred from the  H${\alpha}$ line profiles observed in Be/X-ray binaries, e.g., 4U~0115+63 \citep{Negueruela2001, Negueruela2001A&A}, and are thought to operate within individual cycles of disk dispersal or reconstruction.
However, for GRO~J1008$-$57, the orbital phases of Type I outbursts exhibit a step-like evolutionary features. Except for brief phase jumps occurring at the interval times (tens of days) of the Type II outbursts, the orbital phases remain relatively stable, showing no evidence for a continuous long-term phase drift driven by such density perturbations.  A remarkable stability of Type I outburst phases before the first observed Type II outburst was pointed out by \cite{Kuhnel2013A&A...555A..95K}. 
Therefore, this behavior argues against a global one-armed oscillation-driven phase shift in GRO~J1008$-$57. We note, however, that optical H$\alpha$ observations of GRO~J1008$-$57 show significant variability in line shape, intensity, and asymmetry \citep{Coe2007MNRAS.378.1427C}, consistent with a density perturbed Be disk. Although sparse and predating our X-ray data, these results suggest that global disk perturbations or disk-loss processes may still be present.

Similarly, long-term disk precession \citep{Martin2014ApJ...790L..34M} is expected to generate prolonged phase recovery toward periastron phase, as observed over $\sim500$~d in EXO~2030+375 after its 2021 Type II outburst \citep{Huang2025ApJ...984...66H}. In contrast, no comparable long-duration phase recovery is detected in GRO~J1008$-$57 after Type II outburst, raising questions about the applicability of a precessing Be disk scenario in GRO~J1008$-$57.

In light of the above analysis, the central question therefore emerges: what physical mechanism drives the step-like orbital phase evolution observed in GRO~J1008$-$57? 
After excluding the mechanisms proposed for EXO~2030+375, including the eccentric Be disk driven by KL oscillations, one-armed oscillation and long-term disk precession, the problem  becomes difficult to understand.
Independently of the exact nature of the observed properties, it is clear that Type II outbursts play a key role in restructuring the Be disk.

This is further supported by the oscillatory behavior of the Type I outburst energetics seen in Fig.~\ref{fig:flux}.  
The systematic increase in energy before Type II outbursts suggests gradual mass accumulation in the Be disk, whereas the rapid decline after Type II outbursts indicates temporary depletion of the disk material. The subsequent gradual recovery reflects the rebuilding of the disk density and the mass supply.
Thus, this fluence evolution is naturally explained within a framework of disk depletion and reconstruction cycles associated with Type II outbursts.
Each Type II outburst appears to mark a rapid and different reconfiguration of the disk geometry. 
However, the causal connection between the disk evolution and Type II outbursts remains uncertain. It is plausible that large-scale perturbations or instabilities in the Be disk lead to Type II outbursts, rather than being triggered by them.
Once the Type II outbursts end, the interaction between the neutron star and the Be disk settles into a new temporary configuration, shifting the preferred orbital phase for efficient accretion, and thereby changing the orbital phases of subsequent Type I outbursts.

However, the orbital phases of subsequent Type I outbursts 
change only slightly and quickly transit into a new stable state, as indicated by the small amplitude phase jumps in the step-like evolution (see Fig.~\ref{fig:step_phase_fig} and Table~\ref{tab:step_phase_start}). 
A likely explanation for this observed difference between EXO~2030+375 and GRO~J1008$-$57 is related to the long orbital period of the latter (249.48~d, \citealt{Kuhnel2013A&A...555A..95K}), which imply a lower frequency of neutron star–disk interactions. After disk depletion driven by Type II outbursts, the disk has enough time to rebuild its density and restore its geometric structure 
close to its stable state before Type II outbursts. 
As a result, the orbital phases of subsequent Type I outbursts not only change very slightly compared to before the Type II outburst, but can ultimately still remain stable phases over many orbital periods until the occurrence of next Type II–driven disk reconfiguration event.
Therefore, these processes can naturally explain the observed step-like phase evolution of GRO~J1008$-$57.

In contrast, EXO~2030+375 has a short orbital period ($\sim$46~d; \citealt{Wilson2005ApJ}), implying more frequent neutron star–disk interactions that can drive large-scale, long-term perturbations. As a result, not only Type II outbursts but also sometimes even typical Type I outbursts can induce significant, long-term phase shifts, as observed in 1995 and 2016 \citep{Huang2025ApJ...984...66H,Laplace2017A&A...597A.124L,Wilson2002ApJ}.

Future multiwavelength observations will be crucial for further understanding the Be-disk configuration in GRO~J1008$-$57. In particular, optical spectroscopic monitoring of the H${\alpha}$ line profiles can trace the density structure and dynamical evolution of the Be disk, while polarimetric observations may help constrain the disk geometry and inclination. Combined with continued X-ray monitoring, these observations will provide deeper insight into the disk perturbations and reconfiguration processes associated with Type II outbursts.

\begin{acknowledgements}
This work is based on observations obtained with Swift/BAT and MAXI/GSC. We gratefully acknowledge the teams of both missions for maintaining the instruments, providing high-quality data, and making these data publicly available, which were essential for the completion of this study. This work was supported by the Research Council of Finland  Centre of Excellence in Neutron-Star Physics (project 374064). 
HX acknowledges support from the China Scholarship Council (CSC). 
LJ acknowledges support from the National Natural Science Foundation of China under grant Nos. 12173103 and 12261141691.
\end{acknowledgements}

\bibliographystyle{aa}
\bibliography{allbib}

\end{document}